\documentclass{article}
\usepackage{amsmath}
\usepackage{txfonts,bm}  
\newtheorem{theorem}{Theorem}[section]

\newtheorem{remark}[theorem]{Remark}
\newcommand{\cn}{{\rm cn}}
\newcommand{\dn}{{\rm dn}}
\newcommand{\sn}{{\rm sn}}

\makeatletter
\@addtoreset{equation}{section}
\renewcommand{\theequation}{\@arabic\c@section.\@arabic\c@equation}
\makeatother
\usepackage{graphicx}	
\begin{document}
\begin{center}
 {\Large Ultradiscretization of solvable one-dimensional chaotic maps}\\[4mm]

\textsc{\large Kenji Kajiwara$^1$, Atsushi Nobe$^2$ and Teruhisa Tsuda$^1$}\\[2mm]
$^1$ Faculty of Mathematics, Kyushu University,\\
6-10-1 Hakozaki, Fukuoka 812-8581, Japan\\[2mm]
$^2$ Graduate School of Engineering Science, Osaka University,\\
1-3 Machikaneyama-cho Toyonaka, Osaka 560-8531, Japan
\end{center}
\begin{abstract}
We consider the ultradiscretization of a solvable one-dimensional chaotic
map which arises from the duplication formula of the elliptic
functions. It is shown that ultradiscrete limit of the map and its
solution yield the tent map and its solution simultaneously. A geometric
interpretation of the dynamics of the tent map is given in terms of
the tropical Jacobian of a certain tropical curve. Generalization to the
maps corresponding to the $m$-th multiplication formula of the elliptic
functions is also discussed.
\end{abstract}




\section{Introduction}
In this article, we consider the following map
\begin{eqnarray}
 z_{n+1}=f(z_n)=\frac{4z_n(1-z_n)(1-k^2z_n)}{(1-k^2z_n^2)^2},\label{Schroder:map}
\end{eqnarray}
which admits the general solution
\begin{equation}
  z_n = \sn^2(2^nu_0;k),\label{Schroder:solution}
\end{equation}
describing the orbit in $[0,1]$. Here $\sn(u;k)$ is Jacobi's sn
function, $0<k<1$ is the modulus, and $u_0$ is an arbitrary
constant. In fact, (\ref{Schroder:map}) can be reduced to the duplication
formula of sn function:
\begin{eqnarray}
  \sn(2u;k)=\frac{2\sn(u;k)~\cn(u;k)~\dn(u;k)}{1-k^2\sn^4(u;k)},\\
\cn^2(u;k)=1-\sn^2(u;k),\quad \dn^2(u;k)=1-k^2\sn^2(u;k),
\end{eqnarray}
where $\cn(u;k)$ and $\dn(u;k)$ are Jacobi's cn and dn functions,
respectively. The map (\ref{Schroder:map}) is a generalization of the
logistic map (or Ulam-von Neumann map):
\begin{equation}
  z_{n+1}=4z_n(1-z_n),\quad z_n = \sin^2(2^nu_0).\label{logistic:map}
\end{equation}
The map (\ref{Schroder:map}) was first considered by
Schr\"oder \cite{Schroder} in 1871, and it has been studied by many
authors \cite{KF:chaos,Kohda-Fujisaki,Umeno:PR,Umeno:RIMS}.  It is now
classified as one of the (flexible) Latt\`es
maps \cite{Milnor:Lattes}. In this article, we call (\ref{Schroder:map})
the \textit{Schr\"oder map}.

It is well-known that the Schr\"oder map is conjugate to the tent map
for $X_n\in[0,1]$
\begin{equation}
 X_{n+1}=T_2(X_n)=1-2\left|X_n-\frac{1}{2}\right|
=\begin{cases}
2X_n& 0\leq X_n\leq \frac{1}{2},\\  
2(1-X_n)& \frac{1}{2}\leq X_n\leq 1. 
\end{cases}
\label{tent:map}
\end{equation}
Namely, we have the relation
\begin{equation}
s\circ f\circ s^{-1}=T_2,\quad s(z) =  \frac{1}{K(k)}\sn^{-1}(\sqrt{z};k),
\end{equation}
where $K(k)$ is the complete elliptic integral of the first kind
\begin{equation}
 K(k) = \int_0^1 \frac{dx}{\sqrt{(1-x^2)(1-k^2x^2)}}.
\end{equation}

The purpose of this article is to establish a new relationship between
the Schr\"oder map and the tent map through a certain limiting procedure
called the \textit{ultradiscretization} \cite{TTMS:ultra}.  The method
of ultradiscretization has achieved a great success in the theory of
integrable systems.  From the integrable difference equations, various
interesting piecewise linear dynamical systems have been constructed
systematically, such as the soliton cellular automata
\cite{HT:uTzitzeica,IMNS:uSG,MSTTT:uToda,NT:uBurgers,TH:Permanent,TM:umKdV,TS:SCA,TTM:undKP,TH:uKdV,YYT:pBBS}
and piecewise linear version of the Quispel-Roberts-Thompson (QRT) maps
\cite{Nobe:uQRT0,QCS:uQRT,TTGOR:uP,Tsuda:QRT}. The resulting
piecewise linear discrete dynamical systems can be expressible in terms
of the max and $\pm$ operations, which we call the ultradiscrete
systems. The key of the method is that one can obtain not only the
equations but also their solutions simultaneously. It also allows us to understand the underlying mathematical
structures of the ultradiscrete systems
\cite{HHIK:AM_automata,HIK:crystal,Inoue-Takenawa,Iwao-Tokihiro:upToda,KT:upToda,KOSTY:KKR,KS:utheta,KSY:tau_combinatorialBethe,KTT:Bethe,Nobe:uQRT,TM:bb_and_Riemann}.

In this article, we apply the ultradiscretization to the Schr\"oder map
(\ref{Schroder:map}) and its elliptic solution (\ref{Schroder:solution}).
As a result they are reduced to the tent map and its solution.  We also
clarify the tropical geometric nature of the tent map; we show that the
tent map can be regarded as the duplication map on the Jacobian of a
certain tropical curve.

\section{Ultradiscretization of the Schr\"oder map}
\begin{figure}[ht]
 \begin{center}
\includegraphics[scale=0.5]{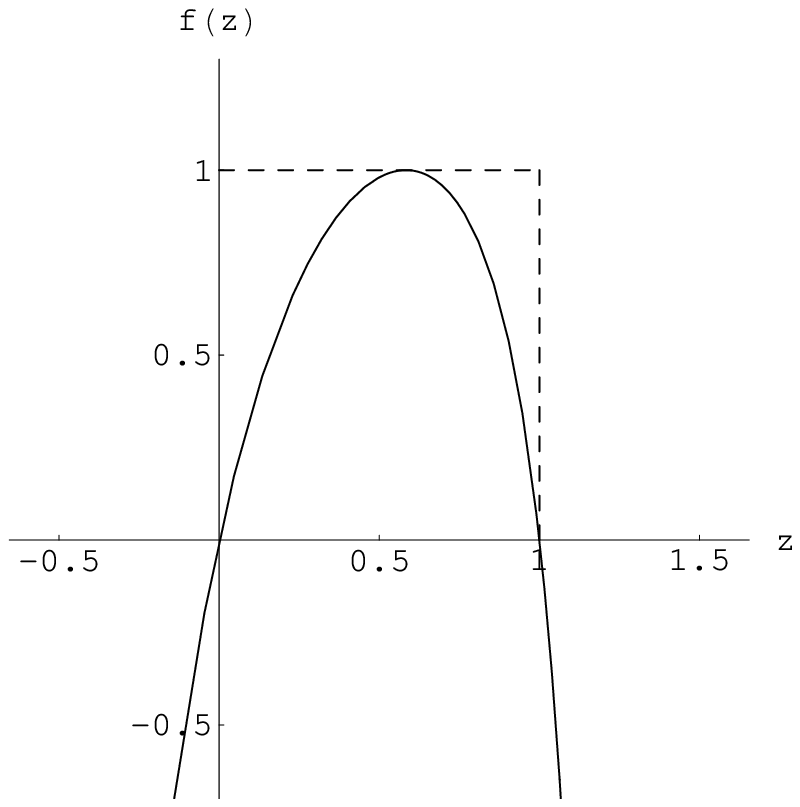}\quad
\includegraphics[scale=0.5]{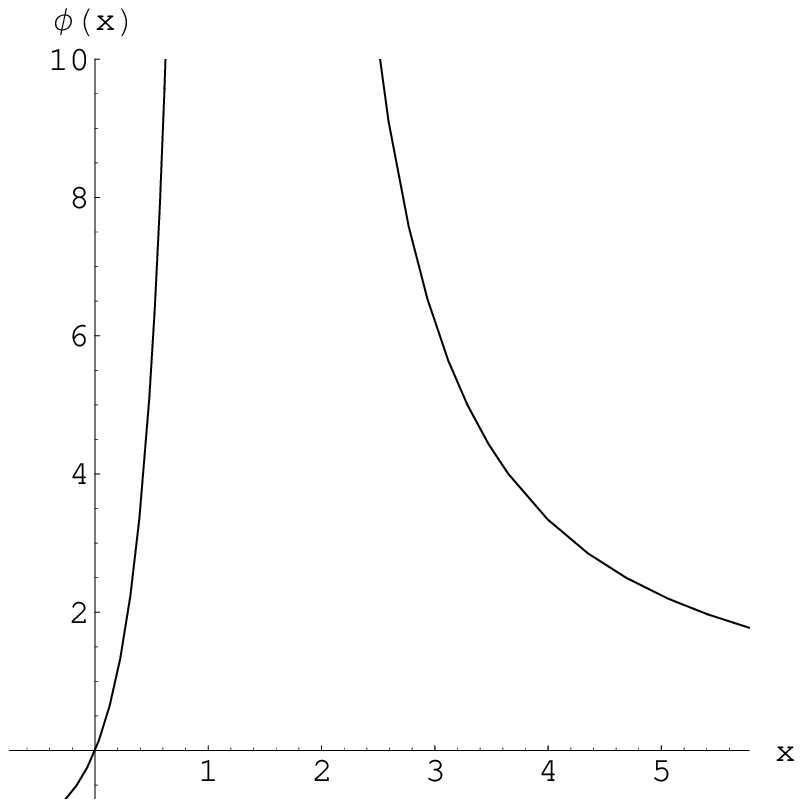}
\end{center}
\caption{Map functions of (\ref{Schroder:map}) (left: $k=0.7$) and
 (\ref{Schroder_tr:map})(right: $k'=0.8$)}\label{fig:map_Schroder}
\end{figure}

The key of the ultradiscretization is the following formula:
\begin{equation}
 \lim_{\epsilon\to +0}
  \epsilon\log\left(e^{\frac{A}{\epsilon}}+e^{\frac{B}{\epsilon}}+\cdots\right)
=\max(A,B,\cdots),\label{eqn:lim_ultra}
\end{equation}
where the terms in $\log$ must be positive, and the dominant term survives
under the limit. We note that the orbit of the map (\ref{Schroder:map}) is always restricted in $[0,1]$ if
the initial value is in this interval. Since this is somewhat too restrictive
for ultradiscretization, we apply the fractional linear transformation
\begin{equation}
 z_n \longmapsto x_n = \frac{z_n}{1-z_n},\label{linear_transformation}
\end{equation}
which maps $[0,1]\rightarrow [0,\infty)$. Then the Schr\"oder map
(\ref{Schroder:map}) and its solution (\ref{Schroder:solution}) are rewritten as
\begin{eqnarray}
x_{n+1}&=&\phi(x_n)=\frac{4x_n(1+x_n)\left(1+k'^2 x_n\right)}{\left(1-k'^2x_n^2\right)^2},
\quad k'^2=1-k^2,\label{Schroder_tr:map}\\
x_n &=& \frac{z_n}{1-z_n}=\frac{\sn^2(2^nu_0;k)}{1-\sn^2(2^nu_0;k)}
=\frac{\sn^2(2^nu_0;k)}{\cn^2(2^nu_0;k)},\label{Schroder_tr:sol}
\end{eqnarray}
respectively. We note that the map (\ref{Schroder_tr:map}) can be
obtained from (\ref{Schroder:map}) by replacing as $z_n\longrightarrow
-x_n$, $k\longrightarrow k'=\sqrt{1-k^2}$. On the level of solution,
this corresponds to Jacobi's imaginary transformation
\begin{equation}
  -i~\sn(iu;k')=\frac{\sn(u;k)}{\cn(u;k)}.
\end{equation}
Figure \ref{fig:map_Schroder} shows the map functions of (\ref{Schroder:map}) and
 (\ref{Schroder_tr:map}). Note that $f(z)$ and $\phi(x)$ have poles at
 $z=\pm 1/k$ and $x=\pm 1/k'$, respectively.

Now we put
\begin{equation}
  x_n = \exp\left[\frac{X_n}{\epsilon}\right],\quad
  k'=\exp\left[-\frac{L}{2\epsilon}\right],\quad
  (0<k'<1,\ L>0).
\end{equation}
Then (\ref{Schroder_tr:map}) is rewritten as
\begin{equation}
 X_{n+1}=F_\epsilon(X_n)=
\epsilon\log\left[\frac{4e^{\frac{X_n}{\epsilon}}(1+e^{\frac{X_n}{\epsilon}})(1+e^{\frac{X_n-L}{\epsilon}})}
{(1-e^{\frac{2X_n-L}{\epsilon}})^2}\right].
\end{equation}
Taking the limit $\epsilon\to +0$ by using the formula
(\ref{eqn:lim_ultra}), we obtain
\begin{eqnarray}
 X_{n+1}&=&F(X_n)=X_n+\max(0,X_n)+\max(0,X_n-L)-2\max(0,2X_n-L)\nonumber\\[1mm]
&=&
\begin{cases}
X_n & X_n<0,\\
2X_n & 0\leq X_n<\frac{L}{2},\\
-2X_n+2L & \frac{L}{2}\leq X_n<L,\\
-X_n+L& L\leq X_n.
\end{cases}
\label{uSchroder:map}
\end{eqnarray}
\begin{figure}[ht]
 \begin{center}
\includegraphics[scale=0.5]{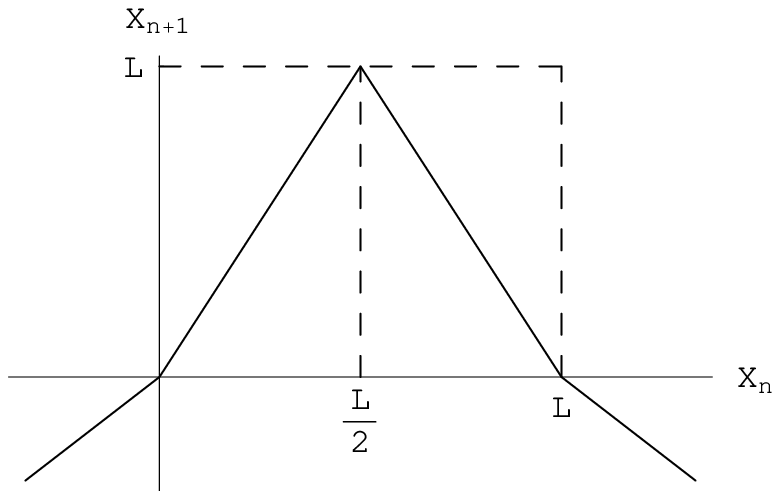}\quad
\includegraphics[scale=0.5]{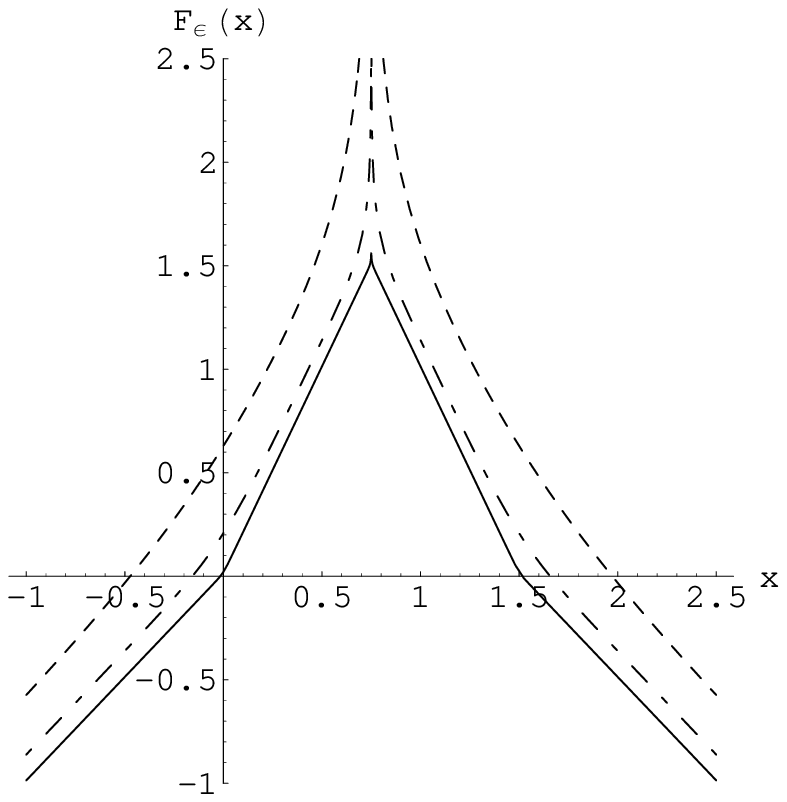}
 \end{center}
\caption{Left: map function of the ultradiscrete Schr\"oder map
 (\ref{uSchroder:map}). Right: limit transition of the map function 
$F_\epsilon(X)$ for $L=1.5$. Dashed line:
 $\epsilon=0.3$, dot-dashed line: $\epsilon=0.1$, solid line:
 $\epsilon=0.01$. }\label{fig:uSchroder_map}
\end{figure}
\begin{remark}
Although the terms in $\log$ in the formula (\ref{eqn:lim_ultra}) must
be positive in general, the negative terms can also exist as long as
they are not dominant in the limit. For example, we have
\begin{equation}
  \lim_{\epsilon\to +0}
  \epsilon\log\left(e^{\frac{A}{\epsilon}}-e^{\frac{B}{\epsilon}}\right)^2
=  \lim_{\epsilon\to +0}
  \epsilon\log\left(e^{\frac{2A}{\epsilon}}-2e^{\frac{A+B}{\epsilon}}+e^{\frac{2B}{\epsilon}}\right)
=2\max(A,B).
\end{equation} 
\end{remark}

We call the map (\ref{uSchroder:map}) the \textit{ultradiscrete Schr\"oder
map}. Figure \ref{fig:uSchroder_map} shows the map function of
(\ref{uSchroder:map}) and limit transition of the function $F_\epsilon(X)$.
The dynamics of the map (\ref{uSchroder:map}) is described as
follows: if the initial value $X_0$ is in $[0,L]$, the map is the tent
map and $X_n\in[0,L]$ for all $n$. If $X_0\in(-\infty,0]$, then
$X_n=X_0$ for all $n\geq 1$. Finally if $X_0\in [L,\infty)$, then $X_1=-X_0+L<0$ and
$X_n=X_1$ for all $n\geq 1$. Therefore the ultradiscrete Schr\"oder map (\ref{uSchroder:map})
is essentially the tent map on $[0,L]$
\begin{equation}
 X_{n+1}=L \left(1 - 2\left|\frac{X_n}{L}-\frac{1}{2}\right|\right),
\quad X_n\in [0,L],\label{uSchroder_tent:map}
\end{equation}
and otherwise the dynamics is trivial. 

Now let us consider the limit of the solution by using the
ultradiscretization of the elliptic theta functions \cite{TTGOR:uP}(see also \cite{KS:utheta,Nobe:utheta,Nobe:uQRT}).
Jacobi's elliptic functions are expressed in terms of the elliptic theta functions
$\vartheta_i(\nu)$ ($i=0,1,2,3$) as
\begin{equation}
 \sn(u;k)=\frac{\vartheta_3(0)\vartheta_1(\nu)}{\vartheta_2(0)\vartheta_0(\nu)},\quad
 \cn(u;k)=\frac{\vartheta_0(0)\vartheta_2(\nu)}{\vartheta_2(0)\vartheta_0(\nu)},
\end{equation}
\begin{equation}
 u=\pi(\vartheta_3(0))^2\nu,\quad k^2=\left(\frac{\vartheta_2(0)}{\vartheta_3(0)}\right)^4,
\end{equation}
where 
\begin{eqnarray}
 \vartheta_0(\nu)&=&\sum_{n\in\mathbb{Z}} (-1)^n q^{n^2}z^{2n},\\
 \vartheta_1(\nu)&=&i\sum_{n\in\mathbb{Z}} (-1)^n q^{(n-1/2)^2}z^{2n-1},\\
 \vartheta_2(\nu)&=&\sum_{n\in\mathbb{Z}} q^{(n-1/2)^2}z^{2n-1},\\
 \vartheta_3(\nu)&=&\sum_{n\in\mathbb{Z}} q^{n^2}z^{2n},
\end{eqnarray}
and $z=\exp[i\pi \nu]$. We parametrize the nome $q$ as
\begin{equation}
  q=\exp\left[-\frac{\epsilon\pi^2}{\theta}\right], \quad\theta>0.
\label{nome}
\end{equation}
Applying Jacobi's imaginary transformation (or Poisson's summation
formula) the elliptic theta functions are rewritten as
\begin{eqnarray}
 \vartheta_0(\nu)&=&\sqrt{\frac{\theta}{\epsilon\pi}}
\sum_{n\in\mathbb{Z}}\exp\left[-\frac{\theta}{\epsilon}\left\{\nu-\left(n+\frac{1}{2}\right)\right\}^2\right],\\
 \vartheta_1(\nu)&=&\sqrt{\frac{\theta}{\epsilon\pi}}
\sum_{n\in\mathbb{Z}}(-1)^n\exp\left[-\frac{\theta}{\epsilon}\left\{\nu-\left(n+\frac{1}{2}\right)\right\}^2\right],\\
 \vartheta_2(\nu)&=&\sqrt{\frac{\theta}{\epsilon\pi}}
\sum_{n\in\mathbb{Z}}(-1)^n\exp\left[-\frac{\theta}{\epsilon}\left(\nu-n\right)^2\right],\\
 \vartheta_3(\nu)&=&\sqrt{\frac{\theta}{\epsilon\pi}}
\sum_{n\in\mathbb{Z}}\exp\left[-\frac{\theta}{\epsilon}\left(\nu-n\right)^2\right].
\end{eqnarray}
Asymptotic behaviour of these functions for $\epsilon\to+0$ is given by
\begin{eqnarray}
 \vartheta_0(0)&\sim&  2\sqrt{\frac{\theta}{\epsilon\pi}}\exp\left[-\frac{\theta}{4\epsilon}\right],\\
 \vartheta_2(0)&\sim&  \sqrt{\frac{\theta}{\epsilon\pi}}\left(1-2\exp\left[-\frac{\theta}{\epsilon}\right]\right),\\
 \vartheta_3(0)&\sim&  \sqrt{\frac{\theta}{\epsilon\pi}}\left(1+2\exp\left[-\frac{\theta}{\epsilon}\right]\right),\\
 (\vartheta_0(\nu))^2&\sim&\frac{\theta}{\epsilon\pi}
\exp\left[-\frac{2\theta}{\epsilon}\left\{((\nu))-\frac{1}{2}\right\}^2\right],\\
 (\vartheta_1(\nu))^2&\sim& \frac{\theta}{\epsilon\pi}
\exp\left[-\frac{2\theta}{\epsilon}\left\{((\nu))-\frac{1}{2}\right\}^2\right],\\
 (\vartheta_2(\nu))^2&\sim& \frac{\theta}{\epsilon\pi}
\left(
\exp\left[-\frac{\theta}{\epsilon}\left\{((\nu))\right\}^2\right]
-\exp\left[-\frac{\theta}{\epsilon}\left\{((\nu))-1\right\}^2\right]
\right)^2,
\end{eqnarray}
where $((\nu))$ is the decimal part of $\nu$, namely,
\begin{equation}
 ((\nu)) = \nu - {\rm Floor}(\nu),\quad 0\leq ((\nu)) < 1.
\end{equation}
Then we have
\begin{eqnarray*}
 k'^2&=&\exp\left[-\frac{L}{\epsilon}\right]=1-k^2=1-\left(\frac{\vartheta_2(0)}{\vartheta_3(0)}\right)^4\sim
\frac{16\exp\left[-\frac{\theta}{\epsilon}\right]\left(1+4\exp\left[-\frac{2\theta}{\epsilon}\right]\right)}
{\left(1+2\exp\left[-\frac{\theta}{\epsilon}\right]\right)^4},\\
x_n&=&\exp\left[\frac{X_n}{\epsilon}\right]=\frac{\sn^2(u;k)}{\cn^2(u;k)}
=\left(\frac{\vartheta_3(0)\vartheta_1(\nu)}{\vartheta_0(0)\vartheta_2(\nu)}\right)^2
\sim\frac{\left(1+2\exp\left[-\frac{\theta}{\epsilon}\right]\right)^2
\exp\left[\frac{2\theta((\nu))}{\epsilon}\right]
}
{4\left(
1 - \exp\left[\frac{\theta}{\epsilon}[2((\nu))-1]\right]
\right)^2},
 \end{eqnarray*}
which yield in the limit $\epsilon\to +0$
\begin{eqnarray}
L&=&\theta,\label{uSchroder:K}\\
X_n 
&=&\theta \left(1 - 2\left|((\nu))-\frac{1}{2}\right|\right),\quad \nu = 2^n\nu_0,\label{uSchroder:sol}
\end{eqnarray}
respectively, where $\nu_0$ is an arbitrary constant. We note that in
taking the limit of $x_n$, we have put the arbitrary constant $u_0$ as
\begin{equation}
 u_0=\frac{\theta}{\epsilon}\nu_0
\end{equation}
so that
\begin{equation}
 \nu=\frac{2^nu_0}{\pi(\vartheta_3(0))^2}=2^n\nu_0~\frac{\frac{\theta}{\epsilon}}{\pi(\vartheta_3(0))^2}
\longrightarrow 2^n\nu_0\quad (\epsilon\to +0).
\end{equation}
One can verify that (\ref{uSchroder:K}) and (\ref{uSchroder:sol}) actually
satisfy the ultradiscrete Schr\"oder map (\ref{uSchroder:map}) or
(\ref{uSchroder_tent:map}) by direct calculation. Therefore we have
shown that through the ultradiscretization the Schr\"oder map
(\ref{Schroder_tr:map}) and its solution (\ref{Schroder_tr:sol}) yield
the map (\ref{uSchroder:map}) (or (\ref{uSchroder_tent:map})) and its
solution (\ref{uSchroder:sol}) simultaneously.

\begin{remark}\hfill
\begin{enumerate}
 \item The fundamental periods of $\frac{\sn^2(u;k)}{\cn^2(u;k)}$ are
       $2K(k)$ and $2iK(k')$.  In the ultradiscretization of the
       elliptic theta functions, we have parametrized
the nome $q$ as (\ref{nome}), which implies that the ratio of half-period
$\tau$ is given by $\tau=i\frac{\epsilon\pi}{\theta}$ and
\begin{equation}
 K(k)=\frac{\pi}{2}(\vartheta_3(0))^2\sim \frac{\theta}{2\epsilon},\quad
 K(k')=-\frac{\pi i}{2}(\vartheta_3(0))^2\tau=\frac{\pi^2 \epsilon}{2\theta}(\vartheta_3(0))^2\sim \frac{\pi}{2},
\end{equation}
as $\epsilon\to +0$. Since we have $u=\frac{\theta}{\epsilon}\nu$, the
       fundamental periods with respect to $\nu$
tend to $1$ and $\frac{i\epsilon\pi}{\theta}$ as
$\epsilon\to +0$. This implies that the ultradiscretization of the
elliptic functions is realized by collapsing the imaginary period and
keeping the real period finite.
 \item The Schr\"oder map (\ref{Schroder:map}) is reduced to the logistic map
(\ref{logistic:map}) for $k=0$. This corresponds to the ultradiscrete Schr\"oder map (\ref{uSchroder:map})
with $L=0$,
\begin{equation}
 X_{n+1}=-|X_n|,
\end{equation}
whose dynamics is trivial, and the solution (\ref{uSchroder:sol}) becomes
$X_n=0$. Therefore ultradiscretization of the logistic map does not yield
an interesting map \cite{K:suurikagaku}. In fact, we see that this case is not consistent
with the ultradiscrete limit, since the asymptotic
behaviour of $K(k)$ and $K(k')$ as $k\to 0$ is given by
\begin{equation}
 K(k)\sim \frac{\pi}{2},\quad K(k')\sim \log\frac{4}{k}.
\end{equation}
\end{enumerate}
\end{remark}

\begin{figure}[ht]
 \begin{center}
\includegraphics[scale=0.5]{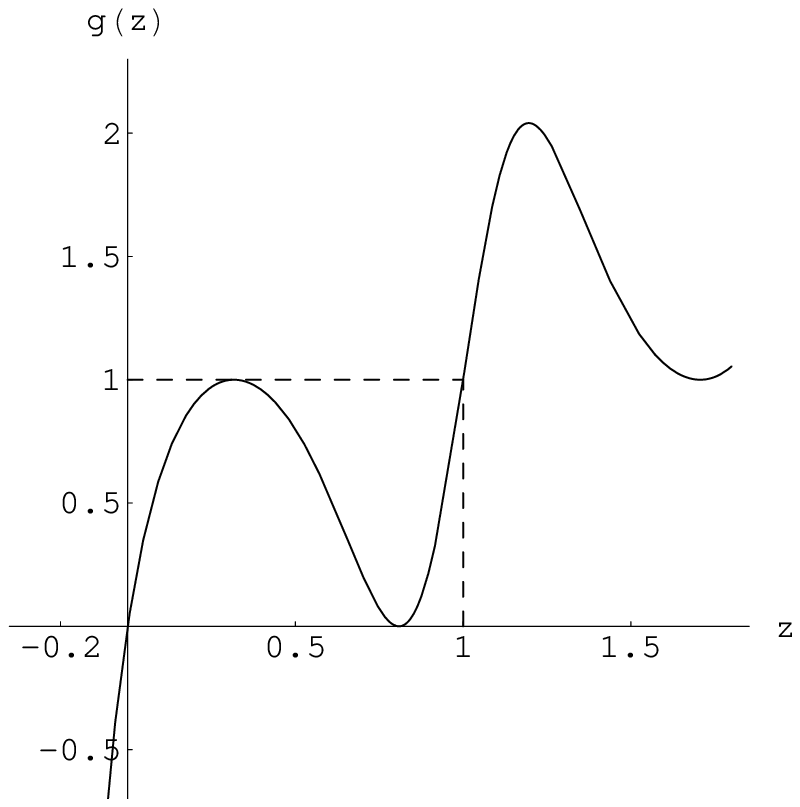}\quad
\includegraphics[scale=0.5]{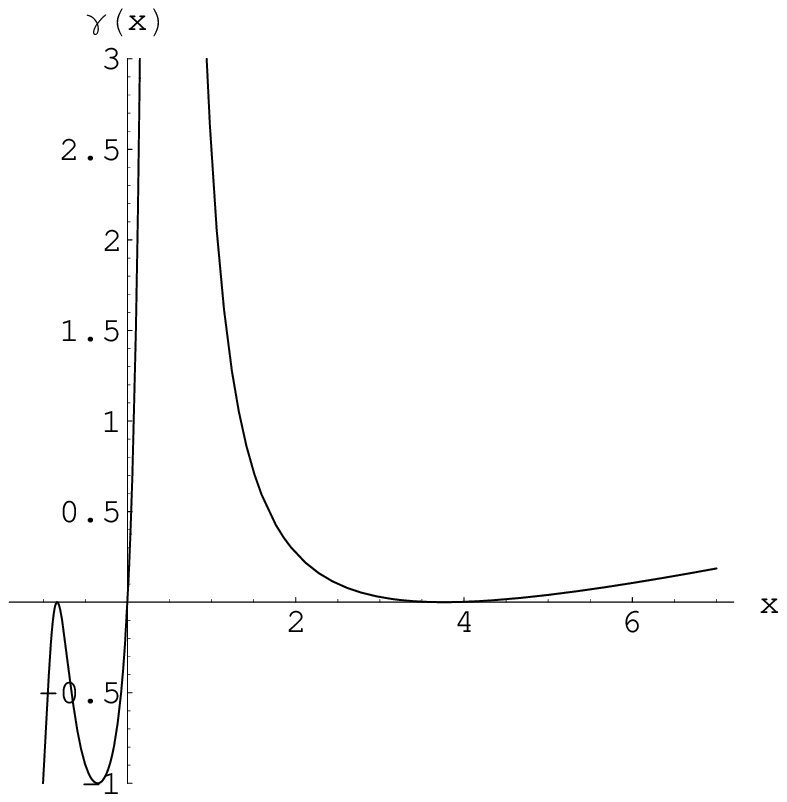}
\end{center}
\caption{Map functions of (\ref{cubic:map}) (left: $k=0.7$) and
 (\ref{cubic_transformed:map})(right: $k'=0.8$)}\label{fig:map_cubic}
\end{figure}

\begin{figure}[ht]
 \begin{center}
\includegraphics[scale=0.5]{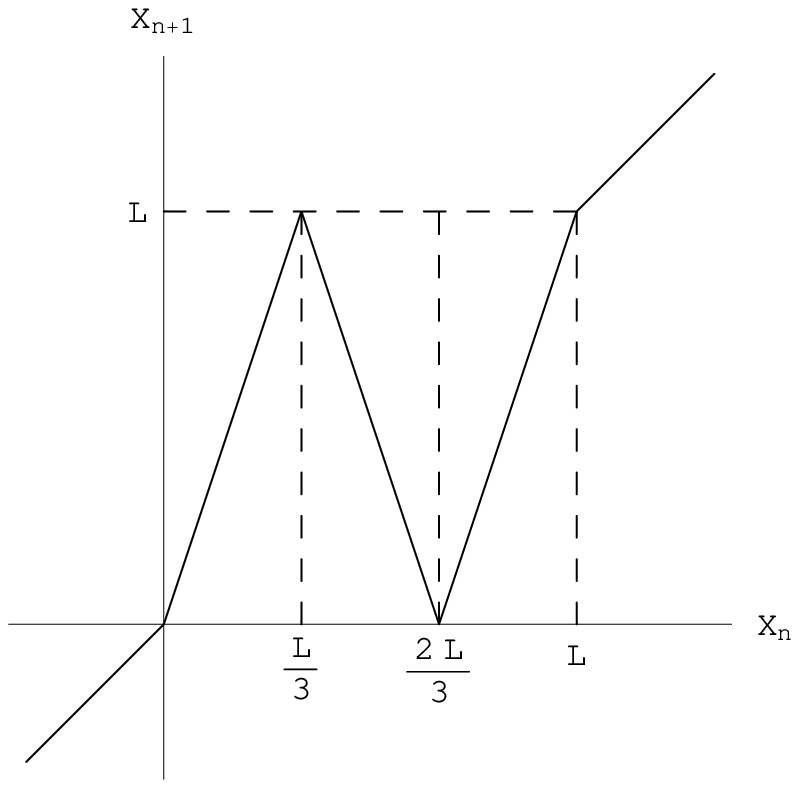}\quad
\includegraphics[scale=0.5]{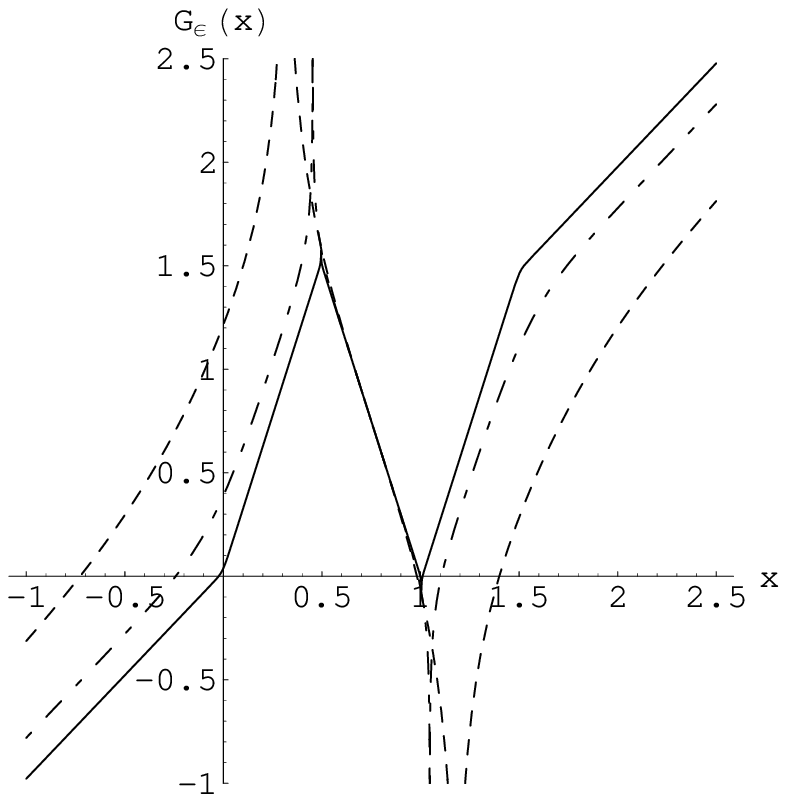}
 \end{center}
\caption{Left: map function of the map
 (\ref{ucubic:map}). Right: limit transition of the map function $G_\epsilon(X)$
for $L=1.5$. Dashed line:
 $\epsilon=0.3$, dot-dashed line: $\epsilon=0.1$, solid line:
 $\epsilon=0.01$. }\label{fig:ucubic_map}
\end{figure}

One can apply the same procedure to the following map which originates
from the triplication formula of $\sn^2$\cite{Kohda-Fujisaki,Milnor:Lattes,Umeno:RIMS}
\begin{equation}
 z_{n+1}=g(z_n)=\frac{z_n\left\{k^4z_n^4-6k^2z_n^2+4(k^2+1)z_n-3\right\}^2}
{\left\{3k^4z_n^4-4k^2(k^2+1)z_n^3+6k^2z_n^2-1\right\}^2},\quad
z_n=\sn^2(3^nu_0;k),\label{cubic:map}
\end{equation}
which is rewritten as 
\begin{equation}
 x_{n+1}=\gamma(x_n)=\frac{x_n\left\{k'^4x_n^4-6k'^2 x_n^2-4(k'^2+1)x_n-3\right\}^2}
{\left\{3k'^4x_n^4+4k'^2(k'^2+1)x_n^3+6k'^2x_n^2-1\right\}^2},\ 
x_n=\frac{\sn^2(3^nu_0;k)}{\cn^2(3^nu_0;k)},\label{cubic_transformed:map}
\end{equation}
by the transformation (\ref{linear_transformation}). The map functions
$g(z)$ and $\gamma(x)$ are illustrated in figure \ref{fig:map_cubic}.
Then ultradiscretization of (\ref{cubic_transformed:map}) yields the
map
\begin{eqnarray}
X_{n+1}=G(X_n)&=&X_n + 2\max(0,X_n,4X_n-2L) - 2\max(0,3X_n-L,4X_n-2L)\nonumber\\
&=&
\begin{cases}
X_n & X_n<0,\\
3X_n & 0\leq X_n<\frac{L}{3},\\
-3X_n+2L & \frac{L}{3}\leq X_n< \frac{2L}{3},\\
3X_n-2L & \frac{2L}{3}\leq X_n < L,\\
X_n& L\leq X_n,
\end{cases}
\label{ucubic:map}
\end{eqnarray}
and its solution
\begin{equation}
  X_n =L \left(1 - 2\left|((\nu))-\frac{1}{2}\right|\right),\quad
\nu=3^n\nu_0.
\end{equation}
Figure \ref{fig:ucubic_map}
shows the map function $G(X_n)$ and the limit transition of the map
function of
\begin{equation}
 X_{n+1}=G_\epsilon(X_n)=\epsilon\log\left[
\frac{e^{\frac{X_n}{\epsilon}}
\left\{e^{\frac{4X_n-2L}{\epsilon}}-6e^{\frac{2X_n-L}{\epsilon}}
-4(e^{-\frac{L}{\epsilon}}+1)e^{\frac{X_n}{\epsilon}}-3\right\}^2}
{\left\{3e^{\frac{4X_n-2L}{\epsilon}}+4(e^{-\frac{2L}{\epsilon}}+e^{-\frac{L}{\epsilon}}))e^{\frac{3X_n}{\epsilon}}
+6e^{\frac{2X_n-L}{\epsilon}}-1\right\}^2}
\right].
\end{equation}
We note that one can directly ultradiscretize the map (\ref{cubic:map})
to obtain (\ref{ucubic:map}), however, the solution $x_n=\sn^2(3^nu_0;k)$
degenerates to the trivial solution $X_{n}=0$. Thus it is important to consider
(\ref{cubic_transformed:map}) in order to obtain the limit which is
consistent with the solution.

It is possible to apply ultradiscretization to the maps arising
from the $m$-th multiplication formula of $\sn^2$
\cite{Kohda-Fujisaki,Milnor:Lattes} in a similar manner.
\section{Geometric description in terms of the tropical geometry}
It is shown in \cite{Inoue-Takenawa,Nobe:uQRT} that the tropical
geometry provides a geometric framework for the description 
of the ultradiscrete integrable systems. Therefore it may be natural to
expect that a similar framework also works well for our
case. In this section, we show that the ultradiscrete Schr\"oder map can
be interpreted as the duplication map on the Jacobian of a certain
tropical curve. As for the basic notions of the tropical geometry,
we refer to \cite{Gathmann,IMS:tropical_book,RST:1st_step}.

We first consider the elliptic curve 
\begin{equation}
 \left[xy-b(x+y)+c\right]^2=4d^2xy, \label{elliptic_curve}
\end{equation}
parametrized by
\begin{equation}
 (x,y)= \left(\frac{\sn^2(u;k)}{\cn^2(u;k)},\frac{\sn^2(u+\eta;k)}{\cn^2(u+\eta;k)}\right),
\end{equation}
where $\eta$ is a constant and $a$, $b$, $d$ are given by
\begin{equation}
b =  \frac{1}{k'^2}\frac{\cn^2(\eta;k)}{\sn^2(\eta;k)},\quad 
c=\frac{1}{k'^2},\quad 
d =  -\frac{1}{k'^2}\frac{\dn(\eta;k)}{\sn^2(\eta;k)},\label{elliptic1_par}
\end{equation}
respectively. Eliminating $\eta$ in (\ref{elliptic1_par}), we see that $b$ and
$d$ satisfy the relation
\begin{equation}
 k'^2d^2=(1+k'^2b)(1+b).\label{bd_relation}
\end{equation}
We may regard the Schr\"oder map (\ref{Schroder_tr:map}) as the
projection of the dynamics of the point on the elliptic curve
(\ref{elliptic_curve}) to the $x$-axis. 

\begin{figure}[ht]
 \begin{center}
\includegraphics[scale=0.3]{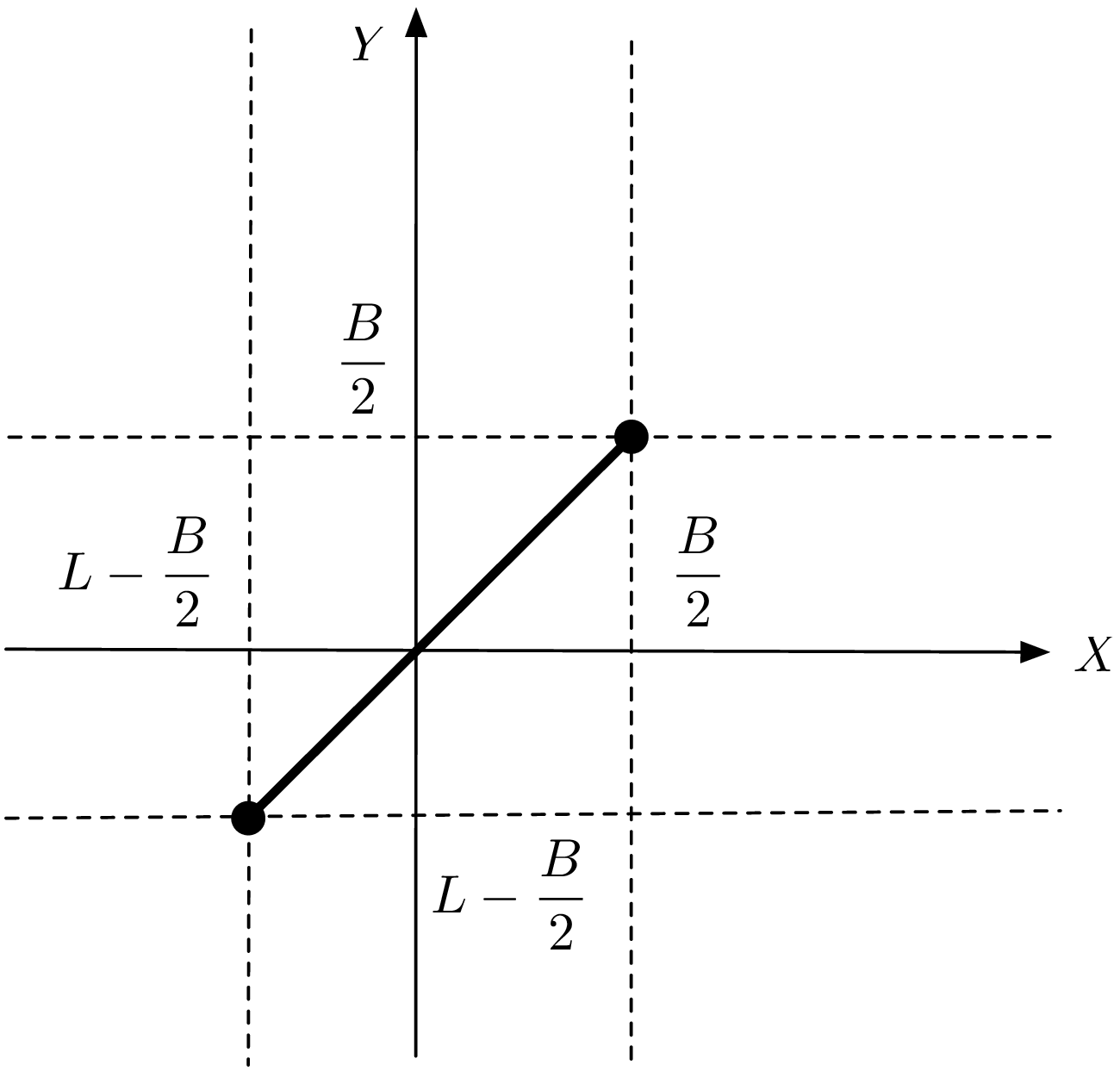}\quad
\includegraphics[scale=0.3]{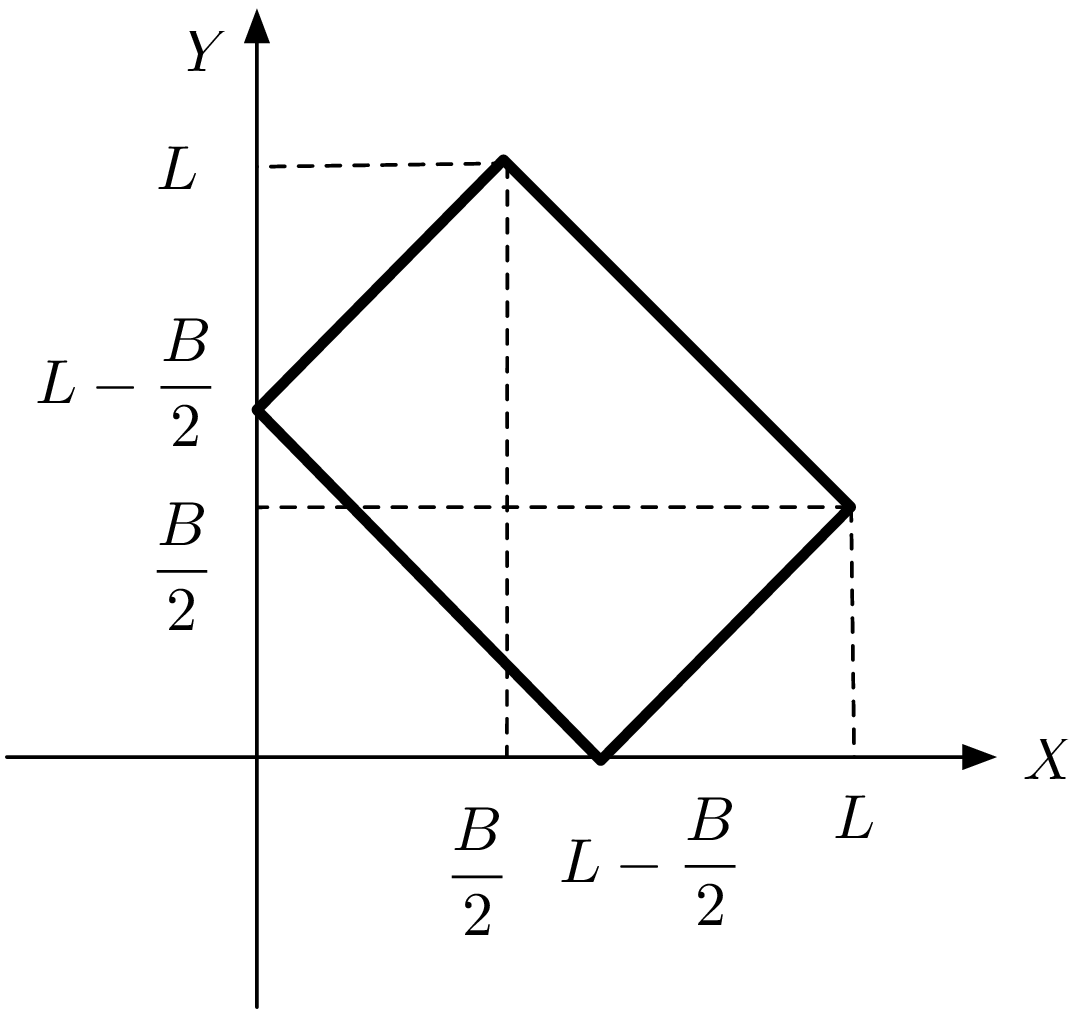}\quad
\includegraphics[scale=0.3]{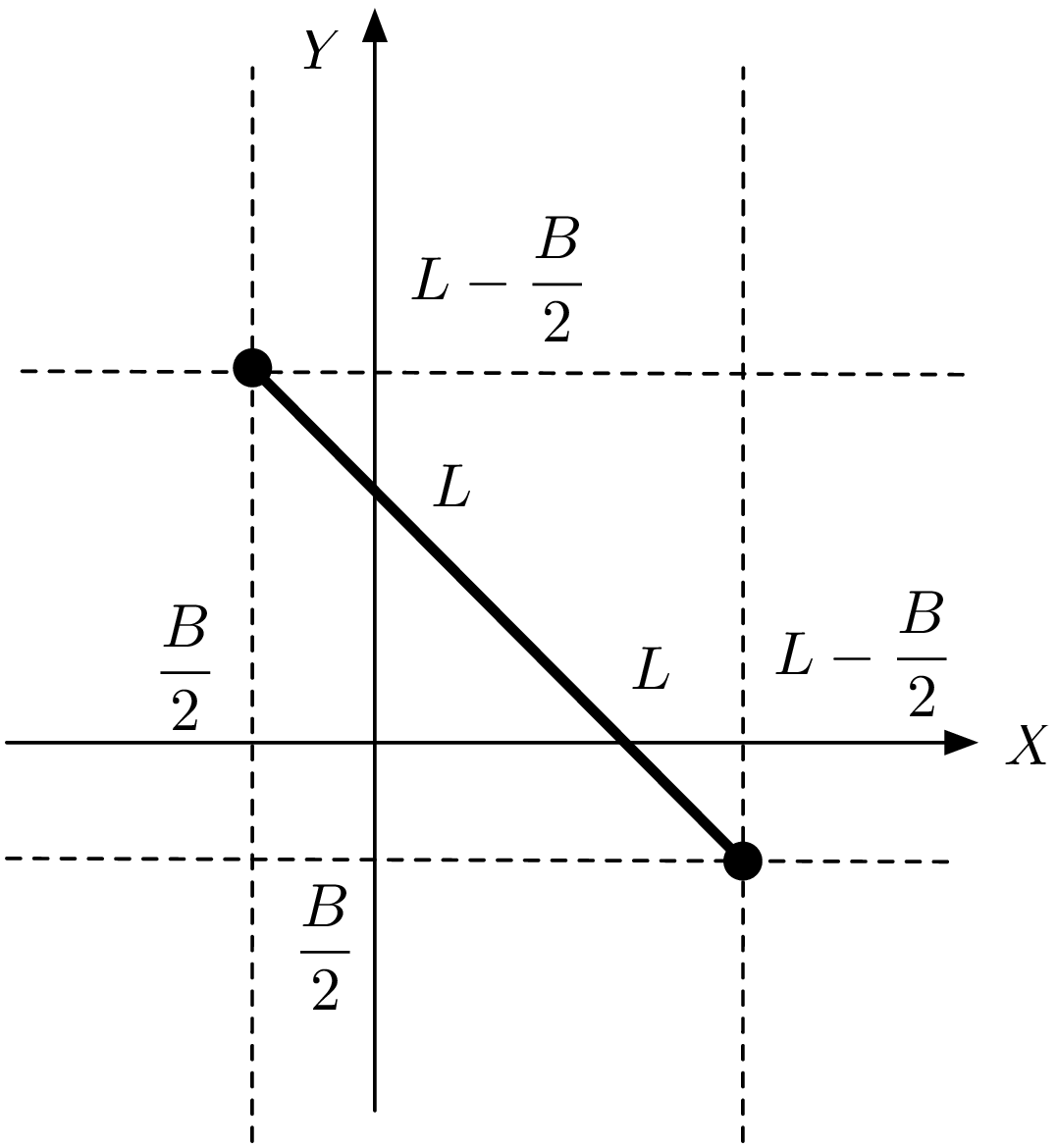}
 \end{center}
\caption{Ultradiscretization of the elliptic curve
 (\ref{elliptic_curve}). Left: case (i), center: case (ii), right: case (iii).}\label{fig:u_elliptic_curve}
\end{figure}

We next apply the ultradiscretization to the elliptic curve. Putting
\begin{equation}
x=e^{\frac{X}{\epsilon}},\  y=e^{\frac{Y}{\epsilon}},\ 
b= e^{\frac{B}{2\epsilon}},\   4d^2=e^{\frac{D}{\epsilon}},\ 
k'=e^{-\frac{L}{2\epsilon}},\ 
c=\frac{1}{k'^2}=e^{\frac{L}{\epsilon}}, \  L>0,
\end{equation}
and taking the limit $\epsilon\to +0$, (\ref{elliptic_curve}) and
(\ref{bd_relation}) yield
\begin{equation}
 \max(2X+2Y,B+2X,B+2Y,2L)=X+Y+D, \label{u_elliptic_curve}
\end{equation}
and
\begin{equation}
 -L+D = \max\left(0,\frac{B}{2}-L\right) +
  \max\left(0,\frac{B}{2}\right), \label{BDL_relation}
\end{equation}
respectively. The condition (\ref{BDL_relation}) gives the following
three cases:
\begin{eqnarray}
\mbox{(i)}&\quad& B>2L>0,\quad D=B, \label{cond1}\\
\mbox{(ii)}&\quad& 2L>B>0, \quad D=L+\frac{B}{2},\label{cond2}\\
\mbox{(iii)}&\quad& 0>B, \quad D=L. \label{cond3}
\end{eqnarray}
For each case, the set of points defined by (\ref{u_elliptic_curve}) is 
(i) a line connecting $(\frac{B}{2},\frac{B}{2})$ and
$(L-\frac{B}{2},L-\frac{B}{2})$, (ii) a rectangle with vertices
$(0,L-\frac{B}{2})$, $(L-\frac{B}{2},0)$, $(L,\frac{B}{2})$ and
$(\frac{B}{2},L)$, (iii) a line connecting $(\frac{B}{2},L-\frac{B}{2})$
and $(L-\frac{B}{2},\frac{B}{2})$, respectively, as illustrated in
figure \ref{fig:u_elliptic_curve}. In the following, we consider only the
case (ii) and we denote the rectangle as $\overline{C}$.

Let us recall some notions of the tropical geometry. The tropical
curve defined by the tropical polynomial
\begin{equation}
 \Xi(X,Y)=\max_{(a_1,a_2)\in{\cal
  A}}(\lambda_{(a_1,a_2)}+a_1X+a_2Y),\quad {\cal A}\in \mathbb{Z}^2,
\end{equation}
is a set of points $(X,Y)\in\mathbb{R}^2$ where $\Xi$ is not smooth.
Here ${\cal A}$ is a finite subset of $\mathbb{Z}^2$ called the
support, and we denote as $\Delta({\cal A})$ the convex hull of ${\cal A}$.
Let $\Gamma_d$ be the triangle in $\mathbb{Z}^2$ with vertices $(0,0)$, $(d,0)$, $(0,d)$. Then the
degree of the tropical curve is $d$ if $\Delta({\cal A})$ is inside
$\Gamma_d$ but not inside $\Gamma_{d-1}$ \cite{Vigeland}. 
The genus of the tropical curve is defined as the first Betti number
of the curve, namely the number of its cycles \cite{Gathmann,Mikhalkin:enumerative,Mikhalkin:application}. 

\begin{figure}[ht]
 \begin{center}
\includegraphics[scale=0.3]{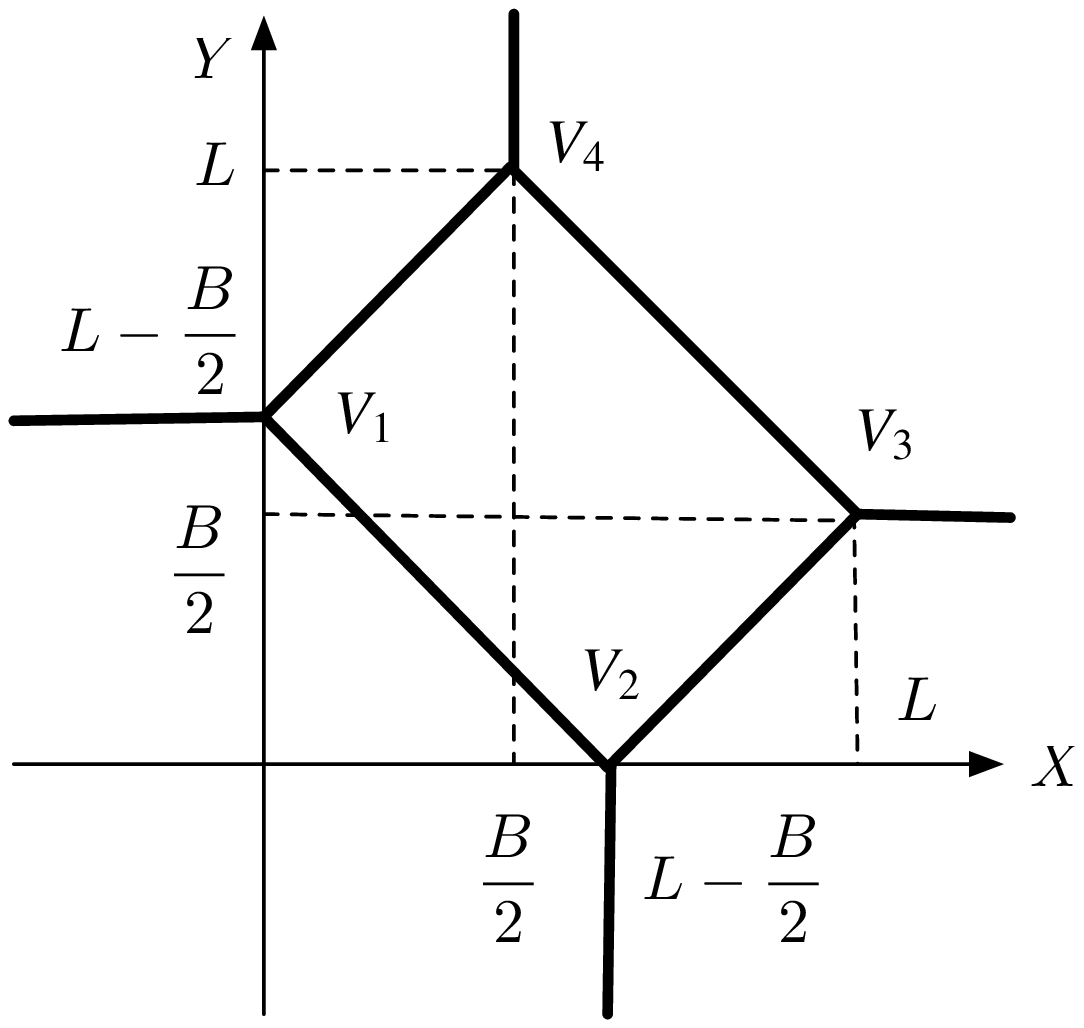}\qquad
\includegraphics[scale=0.35]{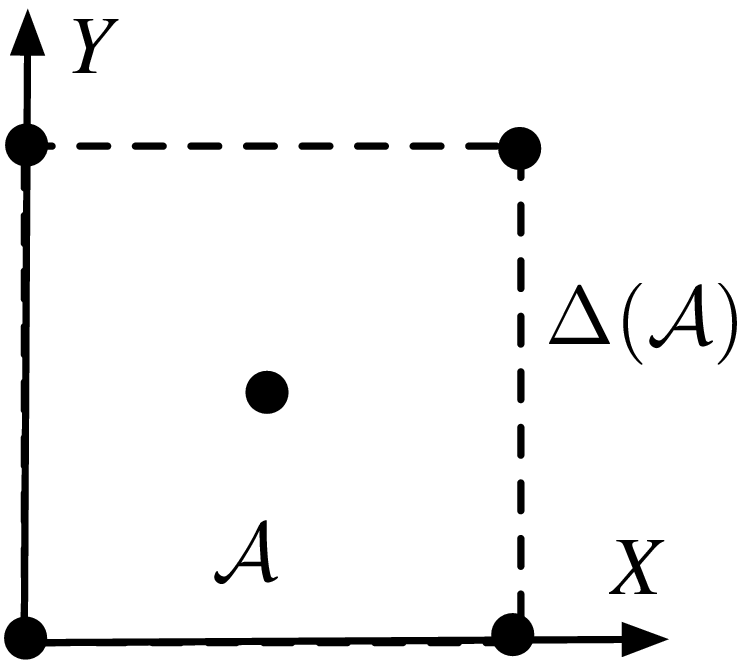}
\caption{Left: tropical curve $C$ defined by
  (\ref{tropical_polynomial}). Right: support of (\ref{tropical_polynomial}).}\label{fig:tropical_elliptic_curve}
\end{center}
\end{figure}

We consider the tropical polynomial
\begin{equation}
\Psi(X,Y)=\max(2X+2Y,B+2X,B+2Y,2L,X+Y+D),\label{tropical_polynomial}
\end{equation}
under the condition (\ref{cond2}). Let $C$ be the tropical curve defined
by $\Psi$, which is illustrated in figure \ref{fig:tropical_elliptic_curve}. 
Then the degree and the genus of $C$ are $4$ and $1$, respectively.
Note that the rectangle $\overline{C}$ is exactly the cycle of $C$.

\begin{figure}[ht]
 \begin{center}
\includegraphics[scale=0.35]{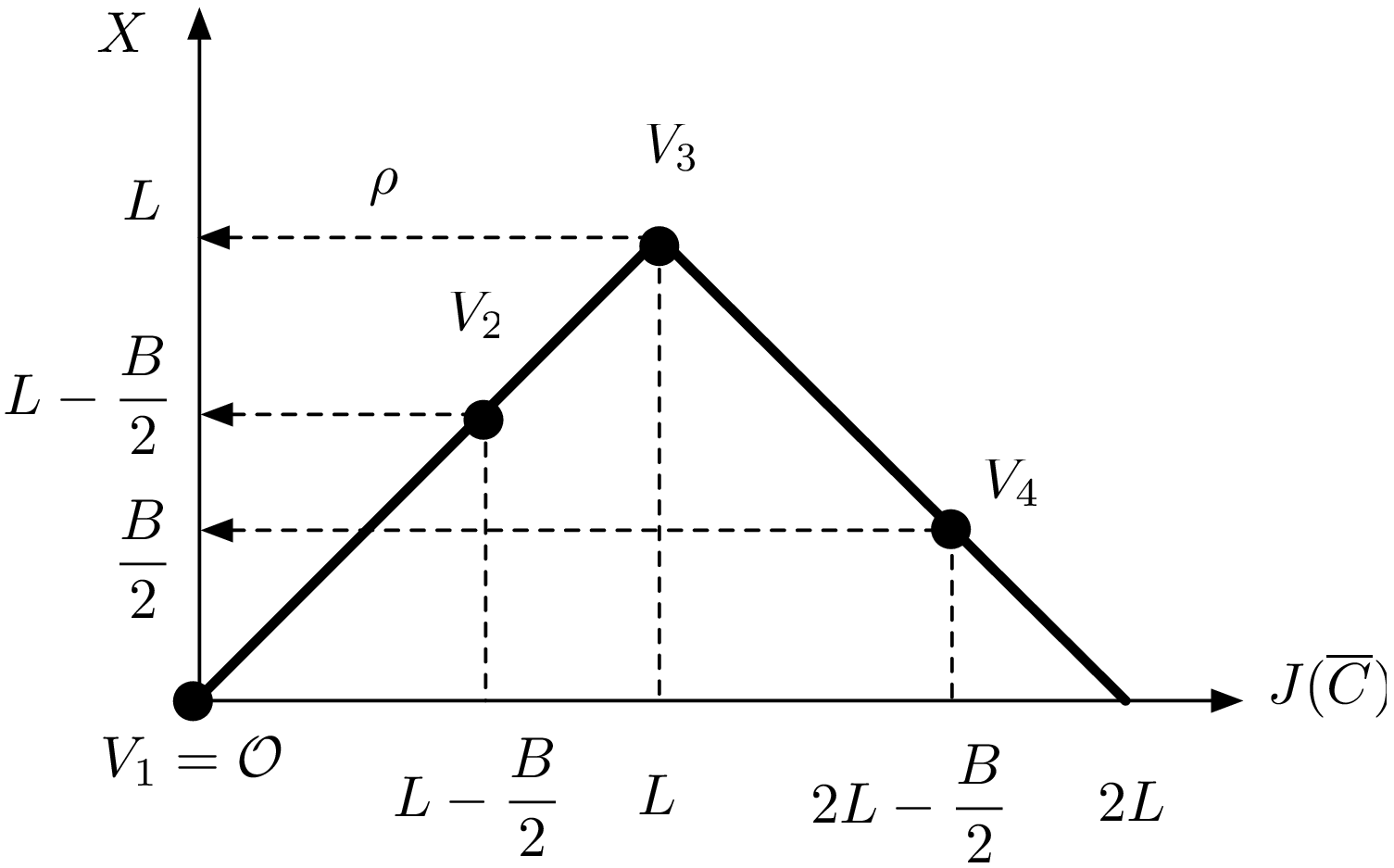}\qquad
\includegraphics[scale=0.35]{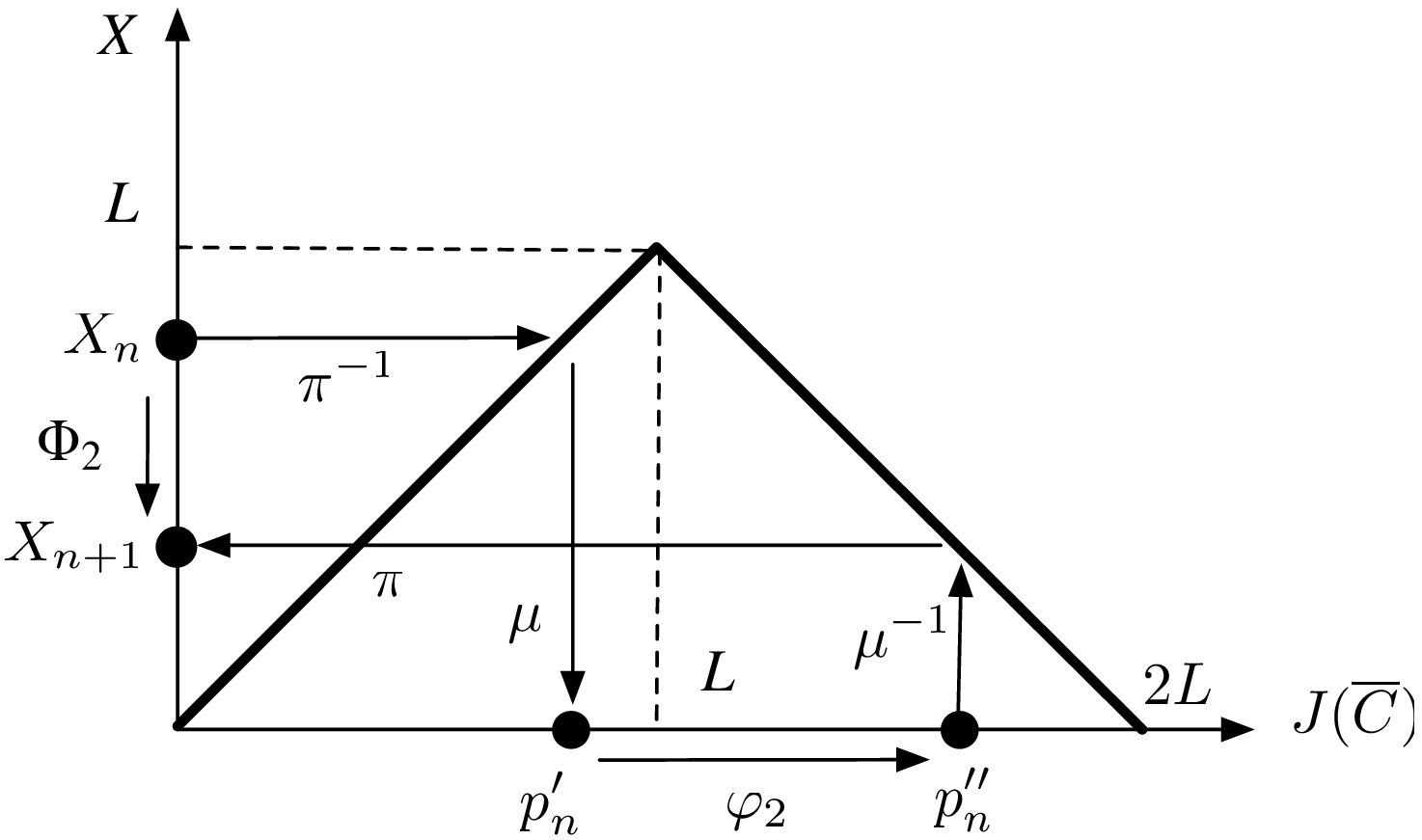}
\caption{Left: correspondence between $X$ and $J(\overline{C})$ by
  $\rho$. Right: duplication map $\varphi_2$ and $\Phi_2$.}\label{fig:tropical_duplication}
\end{center}
\end{figure}

Vigeland \cite{Vigeland} has successfully introduced the group law on the
tropical elliptic curve. Unfortunately, however, his definition of
tropical elliptic curve is limited to ``smooth'' curve of degree 3 and
hence it does not cover our case. Nevertheless, it is possible to define
the tropical Jacobian $J(\overline{C})$ of
$C$ \cite{Inoue-Takenawa,Mikhalkin-Zharkov} and characterize the dynamics
of the ultradiscrete Schr\"oder map (\ref{uSchroder_tent:map}) on it in
the following manner: let $V_i$ and $E_i$ ($i=1,\ldots,4$) be the
vertices and edges of $\overline{C}$ defined by
\begin{equation}
V_1={\cal O}=\left(0,L-\frac{B}{2}\right),\ V_2=\left(L-\frac{B}{2},0\right),\ 
V_3=\left(L,\frac{B}{2}\right),\ V_4=\left(\frac{B}{2},L\right),
\end{equation}
\begin{equation}
 {V_1V_2}=E_1,\quad  {V_2V_3}=E_2,\quad  {V_3V_4}=E_3,\quad  {V_4V_1}=E_4,
\end{equation}
respectively. The length of each edge is given as
\begin{equation}
 |E_1|=\sqrt{2}\left(L-\frac{B}{2}\right),\quad
 |E_2|=\frac{\sqrt{2}}{2}B,\quad
 |E_3|=\sqrt{2}\left(L-\frac{B}{2}\right),\quad
 |E_4|=\frac{\sqrt{2}}{2}B.
\end{equation}
The primitive tangent vector for each edge is 
\begin{equation}
 \bm{v}_1=(1,-1),\quad  \bm{v}_2=(1,1),\quad  \bm{v}_3=(-1,1),\quad  \bm{v}_4=(-1,-1).
\end{equation}
We introduce the total lattice length ${\cal L}$ as the sum of the
length of each edge scaled by the length of corresponding primitive
tangent vector, which is computed as 
\begin{equation}
 {\cal L}=\sum_{i=1}^4\frac{|E_i|}{|\bm{v}_i|}=2L.
\end{equation}
Then the tropical Jacobian $J(\overline{C})$ is defined by
\begin{equation}
 J(\overline{C})=\mathbb{R}/{\cal L}\mathbb{Z}=\mathbb{R}/2L\mathbb{Z}.
\end{equation}
The Abel-Jacobi map $\mu:\ \overline{C}\ \rightarrow\ J(\overline{C})$ is
defined as the piecewise linear map which is linear on each edge satisfying 
\begin{equation}
 \mu(V_1)=0,\ \mu(V_2)=L-\frac{B}{2},\ \mu(V_3)=L,\ \mu(V_4)=2L-\frac{B}{2}.
\end{equation}
Let $\pi:\overline{C}~\rightarrow~\mathbb{R}$ be the projection of the
point on $\overline{C}$ to the $X$-axis. Let $\rho$ be the map defined by
$\rho=\pi\circ\mu^{-1}:\ J(\overline{C})\ \rightarrow\ \mathbb{R}$ which
maps $\mu(P)$ ($P\in \overline{C}$) to the $X$-coordinate of $P$. Here we
note that $\pi^{-1}$ is 1:2 and we define $\pi^{-1}(X)$ to be the point
on $\overline{C}$ whose $Y$-coordinate is smaller.  In this setting,
$\rho(p)$ ($p\in J(\overline{C})$) can be written as
\begin{equation}
 \rho(p)=(\pi\circ \mu^{-1})(p)=
\begin{cases}
 p& 0\leq p\leq L,\\
-p+2L & L\leq p\leq 2L,
\end{cases}
\label{rho}
\end{equation}
as shown in the left of figure \ref{fig:tropical_duplication}.

Now we define the duplication map $\varphi_2: J(\overline{C})\ \rightarrow\
J(\overline{C})$ by
\begin{equation}
 \varphi_2(p)\equiv 2p\ ({\rm mod}~{\cal L}),\quad p\in J(\overline{C}),
\end{equation}
and introduce $\Phi_2:\ \mathbb{R}\ \rightarrow \mathbb{R}$ as the conjugation map
of $\varphi_2$ by $\rho$,
\begin{equation}
 \Phi_2=\rho\circ \varphi_2\circ\rho^{-1} .
\end{equation}
In order to write down the map $\Phi_2$ explicitly, 
we introduce $p',p''\in J(\overline{C})$ for $P=(X,Y)\in \overline{C}$ by
\begin{equation}
p'=\rho^{-1}(X)=(\mu\circ \pi^{-1})(X)=X,\quad
p''=\varphi_2(p')=2p'=2X.
\end{equation}
Then the map $\Phi_2$ is expressed as follows (the right of figure \ref{fig:tropical_duplication}):
\begin{enumerate}
 \item For $0\leq X\leq \frac{L}{2}$ : since $0\leq p''\leq L$,
       (\ref{rho}) implies 
\begin{equation}
 \Phi_2(X)=\rho(p'')=2X.
\end{equation}
 \item For $\frac{L}{2}\leq X\leq L$:  since $L\leq p''\leq 2L$,
       (\ref{rho}) implies
\begin{equation}
 \Phi_2(X)=\rho(p'')=-2X+2L.
\end{equation}
\end{enumerate}
The dynamical system
\begin{equation}
 X_{n+1}=\Phi_2(X_n)=L\left(1-2\left|\frac{X_n}{L}-\frac{1}{2}\right|\right)
=
\left\{
\begin{array}{cl}
2X_n & 0\leq X\leq \frac{L}{2},\\
-2X_n+2L & \frac{L}{2}\leq X\leq L,
 \end{array}
\right.
\end{equation}
coinsides with the ultradiscrete Schr\"oder map
(\ref{uSchroder_tent:map}). Therefore we have shown that the ultradiscrete Schr\"oder map
(\ref{uSchroder_tent:map}) can be regarded as the duplication map on the
Jacobian $J(\overline{C})$ of the tropical curve $C$ defined by the
tropical polynomial (\ref{tropical_polynomial}).

Similarly, we define the triplication map 
$\varphi_3: J(\overline{C})\ \rightarrow\ J(\overline{C})$ by
\begin{equation}
 \varphi_3(p)\equiv 3p\ ({\rm mod}~{\cal L}),\quad p\in J(\overline{C}),
\end{equation}
and introduce $\Phi_3:\ \mathbb{R}\ \rightarrow \mathbb{R}$ as the conjugation map
of $\varphi_3$ by $\rho$,
\begin{equation}
 \Phi_3=\rho\circ \varphi_3\circ\rho^{-1} .
\end{equation}
Then the corresponding dynamical system is given by
\begin{eqnarray}
 X_{n+1}&=&\Phi_3(X_n)=
\begin{cases}
 3X_n& 0\leq X_n\leq \frac{L}{3},\\
-3X_n+2L & \frac{L}{3}\leq X\leq \frac{2L}{3},\\
3X_n-2L & \frac{2L}{3}\leq X\leq L
\end{cases}\nonumber\\
&=& 3X_n - 2\max(0,3X_n-L) + 2\max(0,3X_n-2L),
\end{eqnarray}
which is equivalent to (\ref{ucubic:map}) on $[0,L]$. For general $m$, the $m$-th
multiplication map yields the dynamical system 
\begin{equation}
 X_{n+1}=\Phi_m(X_n)=mX_n + 2\sum_{i=1}^{m-1}(-1)^i \max(0,mX_n-iL),
\end{equation}
which may be regarded as the ultradiscretization of the map 
arising from the $m$-th multiplication formula of $\frac{\sn^2}{\cn^2}$.
\section{Concluding Remarks}
In this article, we have presented a new relationship between two
typical chaotic one-dimensional maps, the Schr\"oder map and the tent
map, through the ultradiscretization. Although the ultradiscretization
has been developed in the theory of integrable systems, the results in
this article imply the possibility of applying the method to wider class
of dynamical systems. Our results also suggest that the tropical
geometry combined with the ultradiscretization provides a powerful tool
to study a piecewise linear map, since the ultradiscretization
translates the geometric background of the original rational map into
that of the corresponding piecewise linear map.  It would be an
interesting problem to study various ultradiscrete or piecewise linear
systems, such as ultradiscrete analogues of Painlev\'e systems, 
generalized QRT maps, and higher-dimensional
solvable chaotic maps in this direction.
\section*{Acknowledgement}
The authors would like to express their sincere thanks to
Prof. Yutaka Ishii for stimulating discussions and valuable information.
This work was supported by the JSPS Grant-in-Aid for Scientific Research
19340039, 19740086, 19840039, and the Grant for Basic Science Research
Projects of the Sumitomo Foundation 071254.

\end{document}